# Unraveling the Role of Morphology on Organic Solar Cell Performance

Biswajit Ray, Pradeep R. Nair and Muhammad A. Alam

School of Electrical and Computer Engineering, Purdue University, West Lafayette, IN 47906

(ray0@purdue.edu, alam@purdue.edu)

**Abstract**

Polymer based organic photovoltaic (OPV) technology offers a relatively inexpensive option for solar energy conversion provided its efficiency increases beyond the current level (6-7%) along with significant improvements in operational lifetime. The critical aspect of such solar cells is the complex morphology of distributed bulk heterojunctions, which plays the central role in the conversion of photo-generated excitons to electron-hole pairs. However, the fabrication conditions for the optimal morphology are still unknown due to the lack of quantitative understanding on the effects of process variables on the cell morphology. In this article, we develop a unique process-device co-simulation framework based on phase-field model for phase separation coupled with self-consistent drift-diffusion transport to quantitatively explore the effects of the process conditions (e.g., annealing temperature, mixing ratio, anneal duration) on the organic solar cell performance. Our results explain experimentally observed trends of open circuit voltage and short circuit current that would otherwise be deemed anomalous from the perspective of conventional solar cells. In addition to providing an optimization framework for OPV technology, our morphology-aware modeling approach is ideally suited for a wide class of problems involving porous materials, block co-polymers, polymer colloids, OLED devices etc.

## *1.* Introduction and Background:

The low-temperature, solution-based, inexpensive manufacture of Polymer based organic solar cell makes it a promising alternative to the classical photovoltaic technologies based on crystalline and amorphous Silicon. Since the low-temperature process is also compatible with flexible light-weight substrates like plastics, organic photovoltaic (OPV) technology reduces installation cost and would conform to novel, non-traditional surfaces[1]. These well known advantages of economic processing , however, are offset by equally well known issues of poor efficiency[2], process sensitivity of short circuit current ($J_{SC}$) and open circuit voltage[3] ($V_{OC}$), and rapid performance degradation at operating conditions[4,5]. Despite a worldwide academic and industrial effort to address these problems, the OPV technology to date has achieved neither the grid parity nor the requisite reliability and a significant improvement in stability and performance is required to ensure commercial viability of this technology.

Performance and lifetime enhancement of the organic solar cells have been conventionally explored by empirical approaches[6,7]. A sound theoretical understanding of the complex process kinetics and their effects on the output current-voltage characteristics of the cell remains poorly addressed in the literature. Thus, it is still unclear from theoretical perspective how the improvement in process conditions, if any, could enhance efficiency or the lifetime of the cell. Therefore, in this manuscript, we develop a comprehensive modeling framework that connects various process variables to the device characteristics so that the performance of the cells is optimized and its ultimate efficiency limit can be achieved. Using our models, (i) we explain some puzzling aspects of the current-voltage (I-V) characteristics of the OPV cell. (ii) We study the effect of annealing conditions on the cell performance and show that there is an optimum anneal duration ($t_{opt}$), uniquely determined by the material parameters and fabrication conditions. (iii) We also find that annealing beyond the optimum duration degrade the output current which has direct implication on the intrinsic reliability of OPV cells.

The manuscript is organized as follows: We first describe the working of OPV and the state of the art modeling approaches. The process modeling approach is described in Sec. 2 followed by detailed transport simulation (for both excitons and charge carriers) in Sec. 3. The process-device simulation framework is then applied (Sec. 4) to study the current-voltage characteristics, process optimization and reliability of the OPV cell.

***1.1 Morphology and Working of OPV:*** The schematic of OPV cell, shown in Fig 1(a), indicates that organic solar cell is morphologically different from the conventional, single-crystalline or thin-film (e.g. a-Si, CIGS, or CdTe) solar cells. The conventional solar cell operates as a vertical PN (or PIN) junction diode with stacked bi-layers of p and n doped regions. However, for OPV cells, the interface between the two interpenetrating polymers (called electron donor (D) and electron acceptor (A), respectively) is neither vertical nor lateral, but it is randomly dispersed throughout the volume of cell (hence it is also called bulk heterojunction (BH) cell). Moreover, unlike classical solar cells, both the donor and the acceptor materials of OPV are connected to the top and bottom electrodes directly, making it necessary that the anode and cathode electrodes have different work functions for the collection of charge carriers in the respective contacts. In addition to difference in work function, additional blocking layers are often used so that the carriers cannot escape through the wrong contact.

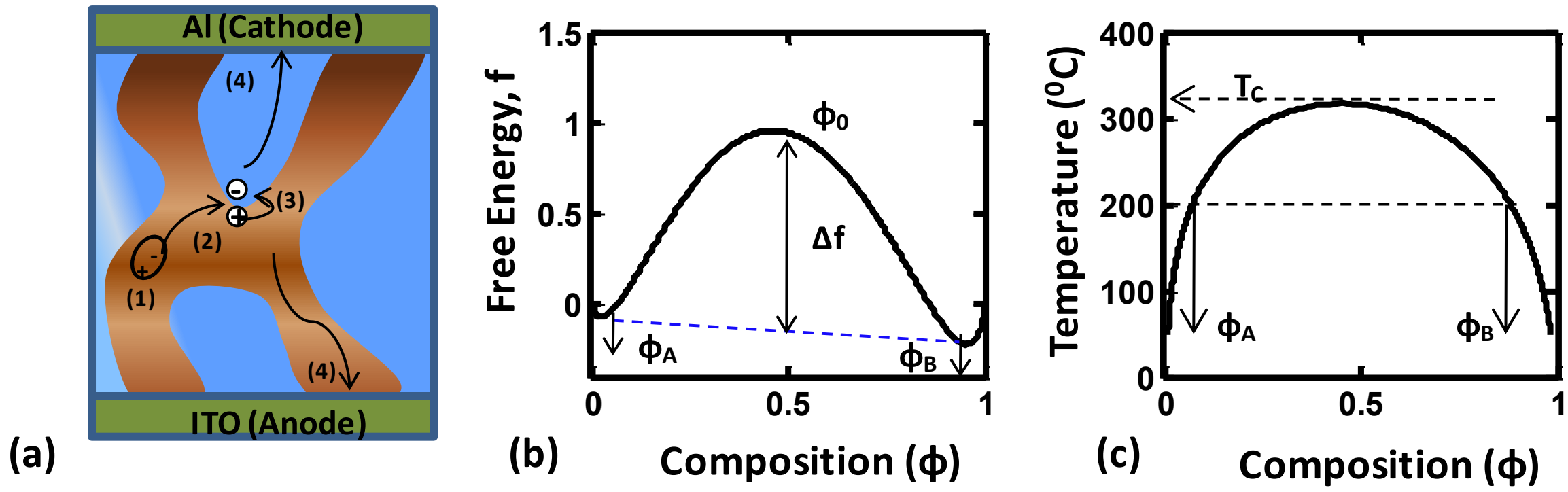

Figure 1: (a) Schematic of donor/acceptor bulk heterojunction solar cell. The light blue region is the acceptor phase (PCBM) and the dark brown region is the donor phase (P3HT). Excitons (denoted by + and - sign), generated in polymer, charge separates at the interface. Charge carriers are denoted by circles with minus sign for electron and plus for hole. (b) Free energy density function for the polymer-fullerene pair is plotted as a function of the composition of the mixture. $\phi = 0$ indicates pure acceptor molecules (PCBM) and $\phi = 1$ indicates donor molecules (P3HT). The minima of the free energy curves ($\phi_A$ and $\phi_B$) denote the equilibrium composition of the phase segregated donor and acceptor phases. The free energy curve implies that the mixture with an initial composition $\phi_0$ separates into two phases (donor and acceptor) with composition $\phi_A$ and $\phi_B$, thus lowering the free energy by $\Delta f$ . (c) The typical phase diagram shown for the polymer fullerene system. The phase diagram dictates the equilibrium composition ($\phi_A$ and $\phi_B$) of the phase-segregated phases for a given temperature. We note that there is a critical temperature $T_C$ beyond which no phase separation is possible. Below $T_C$ phase separation takes place by spinodal decomposition or nucleation and growth depending on initial conc.

To appreciate the importance of complex morphology of OPV cells, let us briefly consider the four sequential processes[8,9] defining the electrical operation of the BH solar cell (Fig. 1a). First, when the photon transmits through the substrate and the electrode (TCO), it is absorbed in the active layer consisting of the D-A polymer blend, generating a strongly bound electron-hole pair ($E_b \sim (0.1-1)\, eV$) called exciton. Next, these excitons diffuse within the disordered active layer morphology defined by respective phases: if they find the donor/acceptor interface within its diffusion length ($L_{ex} = \sqrt{D_{ex}\tau_{ex}} \approx (5-10)\, nm$), they are dissociated into free charge carriers by the quasi-electric field at the heterojunction (step 3 in Fig. 1a); otherwise, the excitons are irreversibly lost to self-recombination with corresponding loss in PV efficiency. Therefore, the distributed donor/acceptor interface is the key innovation of BH-OPV because regardless the origin of an exciton within the active volume, the exciton finds the distributed interface within the diffusion length ($L_{ex}$) so that they may be dissociated into free charge-carriers with very high probability. After exciton dissociation, electrons are transferred to the acceptor material, while holes remain in the donor region. Once electrons and holes are spatially separated, in the fourth and final step, the built-in electric field (created by the difference in work functions of the front and back electrodes) sweeps the free electrons and holes to the respective contacts, and eventually to the load connected to the solar cells (step 4 in Fig. 1a). Note that unlike exciton dissociation, the complexity in the morphology has detrimental effect on the electron/hole collection efficiency. Thus, the performance of the solar cell is dictated by the counterbalancing impact of active layer morphology[10,11] on exciton dissociation and charge transport. Therefore, it has always been a challenge to find the optimum process conditions that define maximum efficiency of the cell.

***1.2 State of the art Modeling Approach:*** A large number of elegant experiments[7,12,13] have been conducted in last few years to deconvolve the effect of the process variables on the ultimate efficiency of the cell. These experiments show that in addition to the choice of constituent polymers[14], the solvent[11,15] in which they are mixed, the mixing ratio of the polymers[16] and the annealing conditions (anneal time and temperature)[13,17-20] , etc have crucial effects on the performance of the cell. The main consequence of the choice of various process variables is that they alter the nature of the underlying active layer morphology in significant ways - a conclusion supported by the various characterization experiments like electron tomography[21,22] of the 3D active layer, small and wide angle XRD[23], TEM images[24], etc. In sum, there exist a large amount of experimental results regarding the effects of various process parameters on the performance of the cell. However, since those parameters are themselves correlated and affect each other in non-trivial ways, the combination of process parameters that optimizes cell performance is generally unknown and has only been explored by empirical approaches.

Similarly from the theoretical side, there have been many interesting papers that focus exclusively on process[25] or device[26] modeling, although the crucial problem of process-device co-modeling remains largely unaddressed. For example, few groups[9,25,27] have explored the process evolution of nano-morphology of organic solar cells by cellular automata approach. While these models demonstrate the complexity of structure and the time evolution of the morphology, the model does not explicitly capture the process conditions, surface tension effect, the nature of the solvent, etc. Due to the lack of proper description of active layer morphology and the complexity involved in solving transport equations on the disordered polymer networks, most of the transport study of OPV is based on effective media theory where active layer is treated as homogenized material. Given the scale of phase segregation in typical OPV samples, this approach clearly cannot lead to morphology aware predictive model. In short, despite the enormous experimental/theoretical literature concerning OPVs, transport in this meso-structured, phase-separated donor/acceptor morphology remains an unsolved problem, which is not readily amenable to classical analysis but inherently requires a new morphology-aware modeling approach, as discussed in this paper.

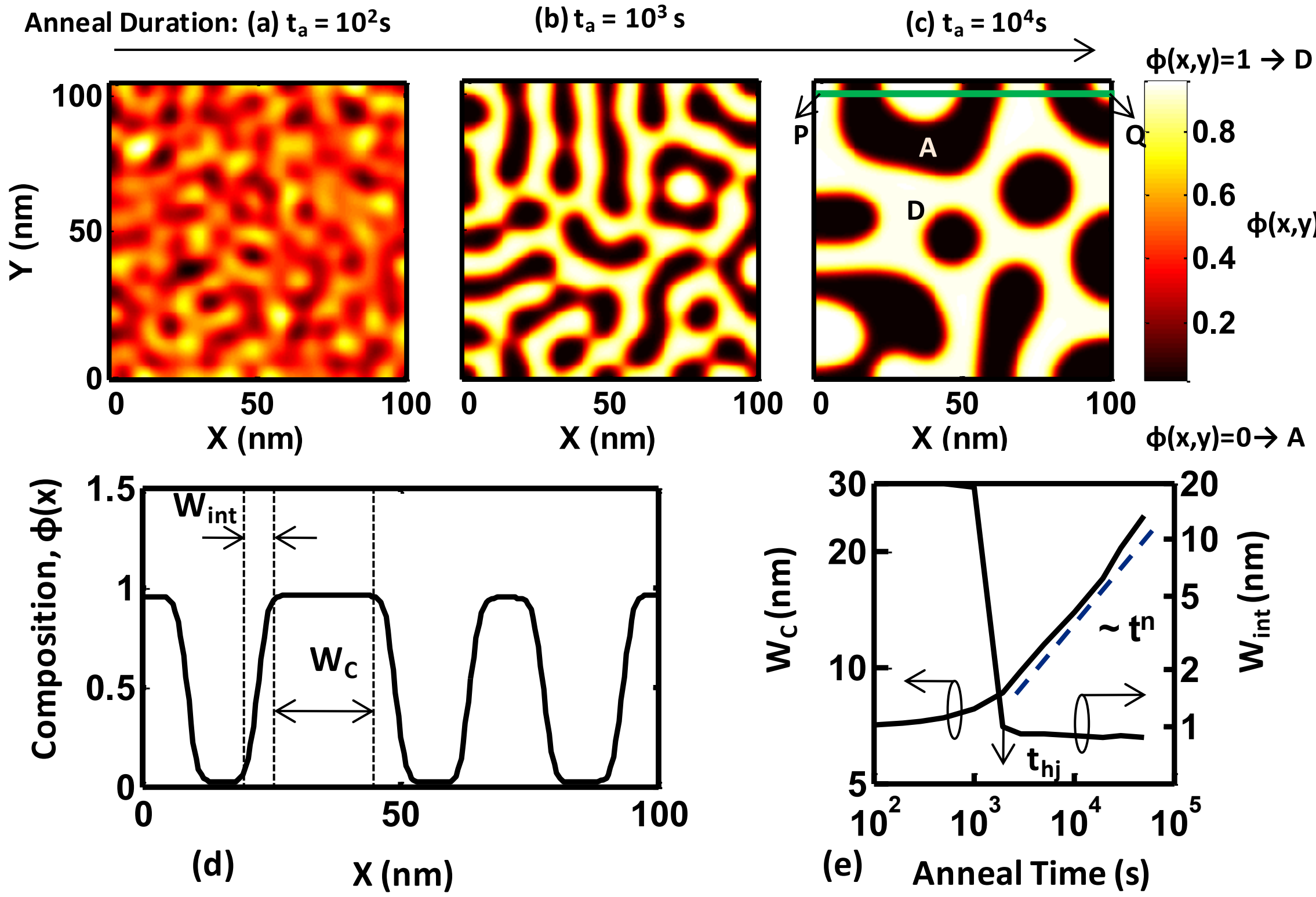

Figure 2: Top view of the morphology of the active layer corresponding to three different anneal time is shown in (a-c). The dark regions in the figure is for acceptor phase ($\phi \sim 0$) and the bright regions is for donor phase ($\phi \sim 1$). (d) The concentration variation along the 1D cut PQ (the green line in Fig. 1c) is plotted. Here $W_{it}$ is the interfacial width and $W_c$ is the average cluster size. (e) Average characteristic cluster width ($W_C$) grows with anneal time ($t_a$) according to a power law as shown in the figure. The evolution of the diffuse interface ($W_{int}$) between the two polymers with anneal time is shown in the RHS axes of the figure. The values of the various parameters used to simulate the morphology are: Anneal temperature, $T$ = $120^0$ C, $\kappa = 10^{-10}$ J/m, $M_0 = 2\text{x}10^{-28}$ $m^5$/J-s and $v_{site} = 5\text{x}10^{-30}$ $m^3$.

## *2.* Process Model:

Spincoating is a popular film forming technique for the fabrication of the polymer solar cell[28]. In this technique, the donor and the acceptor polymers are mixed in the presence of a solvent and the solution is then applied on to a rotating substrate. The film is then thermally annealed so that the solvent is evaporated out and the phase segregated interpenetrating structure emerges within the polymer mixture. The important process variables that affect the shape/structure of this phase segregated nano morphology are anneal temperature, anneal time, initial mixing ratio of the D/A polymer molecules, nature of solvent, etc. Since these process variables dictate ultimate cell performance by controlling the morphology of the phase-segregated polymer blend, it is important to model the morphology as a function of process variables systematically and comprehensively.

*Free Energy Approach to Model Phase Separation:* Although the phase-segregation of polymer blend can be described by cellular-automata approach [9,25,27], an alternate and broadly validated approach of phase separation of the donor and the acceptor polymer is based on the framework of Flory-Huggins mean field theory[29]. According to this theory, phase separation process between a pair of polymers depends on the free energy of mixing ($f$) between the D/A polymers as described by the following equation[30]

$$f = \frac{k_B T}{v_{site}}\left[\frac{\emptyset \ln(\emptyset)}{N_D} + \frac{(1-\emptyset)\ln(1-\emptyset)}{N_A} + \chi\emptyset(1-\emptyset)\right] . \quad (1)$$

The first two terms inside the braces of the above equation correspond to the entropy of mixing, which wants to keep the mixture in the homogeneous/mixed form and the third term is for enthalpy of formation (or the interaction energy), describing the tendency of the mixture to get phase segregated. The local composition (volume fraction of donor) of the mixture is represented by $\phi(x,y,z)$; $N_D$ and $N_A$ are the number of monomer units in the respective donor, acceptor polymer chains (also known as degree of polymerization of D/A polymers), $v_{site}$ is the volume of the reference site in the Flory-Huggins lattice model[29] , and $\chi$ is the interaction parameter between the donor acceptor monomers, also known as Flory mixing parameter. The Flory parameter $(\chi)$ is inversely proportional to temperature $(T)$ and is a very important parameter in free energy function as it essentially decides whether a polymer pair will eventually phase segregate or not (positive $\chi$ favors phase segregation). The polymer chain lengths $N_A$ and $N_B$ introduce asymmetry in the free energy function and eventually dictate the composition of each phase as shown in Fig. 1(b).

The free energy plot in Fig. 1(b) illustrates some very important aspects of the phase-separated morphology. The two minima of the curves (denoted as $\emptyset_D$ and $\emptyset_A$ in Fig. 1(b)) represent the equilibrium composition of donor and acceptor phases, respectively. In other words, after phase segregation, the donor phase is not composed of purely donor polymer molecules, but also contains the donor and acceptor molecules in the ratio defined by $\emptyset_D$. With this compact form of the free energy function (Eq. (1)), we plot the co-existence curve in Fig. 1c, which describes the effect of the anneal temperature on the equilibrium composition of the phase segregated D/A domains. We note the existence of the critical anneal temperature $T_c$ beyond of which equilibrium phase composition is not possible, i.e., the mixture remains in homogeneous state regardless of annealing. The basic free energy function described above can be generalized to include the effects of the substrate strain[31], elasticity or the solvent evaporation[31] etc. by explicitly adding the corresponding terms in the free energy function. For the broad class of BH solar cells, however, this simplest form of free energy (Eq. 1) provides the essential description of the evolution of morphology as a function of process conditions.

*Process Kinetics:* The evolution of the active layer morphology with anneal time depends upon the kinetics of the polymer-polymer phase separation process, which is generally described by the well-known and broadly validated Cahn-Hilliard (C-H) equation[32,33] as written below:

$$\frac{\partial \emptyset}{\partial t_a} = M_0\left(\nabla^2 \frac{\partial f}{\partial \emptyset} - 2\kappa \nabla^4 \emptyset\right) \quad . \qquad (2)$$

C-H equation evaluates the local composition $(\phi(x, y, t))$ change with anneal time based on the minimization of the total free energy. The first term on the right hand side of eq. (2) is a diffusive component with diffusion coefficient depending on the change of free energy. The second term accounts for the surface tension effect due to the formation of the diffused interface. $M_0$ is the effective mobility parameter (dictated by nature of solvent and the anneal temperature T) and $\kappa$ is the gradient energy coefficient. All these parameters can be accurately determined experimentally for a given donor/acceptor polymer pair as described in Ref[31]. Despite the phenomenological nature, the C-H equation with appropriate parameterization captures wide range of phenomena including spinodal decomposition, phase-segregation, phase ordering dynamics, nano-particle dispersion, Oswald ripening, fractionation of polymers with solvent concentration, etc[33].

*Implementation and Validation:* We use spectral method to solve the time dependent C-H equation (Eq. 2) in 3D space (details of numerical implementation is described in literature[34]). Fig. 2(a-c) shows the evolution of the phase segregated BH active layer morphology with anneal time, obtained by numerical solution of CH equation with zero flux boundary conditions. Indeed, Ref.[35] provides a beautiful experimental confirmation of this blend-dependent morphology predicted by the Cahn –Hilliard equation coupled with Flory-Huggins free energy approach.

There are two characteristics features of the nano-morphology defined by the phase segregation process. First, while the geometry of the meso-structure lacks any specific order/shape, it is still characterized by an average domain width, $\langle W_C \rangle$, that increases systematically with anneal time $t_a$ (that eventually leads to Oswald ripening[33]). Second, the interface between the donor and the acceptor phases is a diffuse interface with a finite width. The width of the diffuse interface $(\langle W_{int} \rangle)$ is defined by the region having composition variation between $\phi_A$ and $\phi_D$, and $\langle W_{int} \rangle$ decreases with anneal time. Both $\langle W_C \rangle$ and $\langle W_{int} \rangle$ are crucial in determining the device performance and hence they are explicitly defined in Fig. 2(d) with a 1D cut of the actual morphology. We also find that the growth of $\langle W_C(t_a) \rangle$ with anneal duration follows a power law (refer to Fig. 2(e)) that stabilizes with constant power-exponent within minutes of the initiation of the phase separation and this power law is given by the well known Lifshitz-Slyozov law of phase-segregated polymers, i.e.[30],

$$\langle W_C(t_a) \rangle \sim \left[M_0 f'' \sqrt{\kappa \Delta f}\; t_a\right]^n. \qquad (3)$$

Experimentally, the characteristic length scale of $\langle W_C \rangle$ is interpreted as the inverse of the peak-vector $(q_{max})$ from the light-scattering experiments[30], i.e. $\langle W_C \rangle \sim 2\pi/q_{max}$ and measurements of polymer blends confirm the robustness of the power-law. We reproduce this power-law in Fig. 2(e) to validate the numerical implementation of our model. We will use $W_C(T_{anneal}, t_a, N_A, N_D)$ later to optimize the performance of the solar cell. In addition, we note that the process model is not only limited to BH solar cells, but it is completely general and applicable to any process that involves phase segregation.

## *3.* Device Model

In this section, we will explore the various sub-processes (e.g., optical absorption, exciton transport, charge transfer across the D/A interface, and electron-hole transport in the meso-structured active layer) that translate the photons incident on the solar cell to the current flow at the terminal of the solar cells. A discussion of the sub-processes follows.

### 3.1. Optical absorption

Once the phase-segregated meso-structure is fabricated, and the back contacts are deposited, the first concern is the efficiency of light absorptions of the solar cell. In order to calculate the spatially dependent photon absorption rate, multiple reflections and interference among the electric and magnetic fields need to be taken into account for the multi-layered OPV structure.

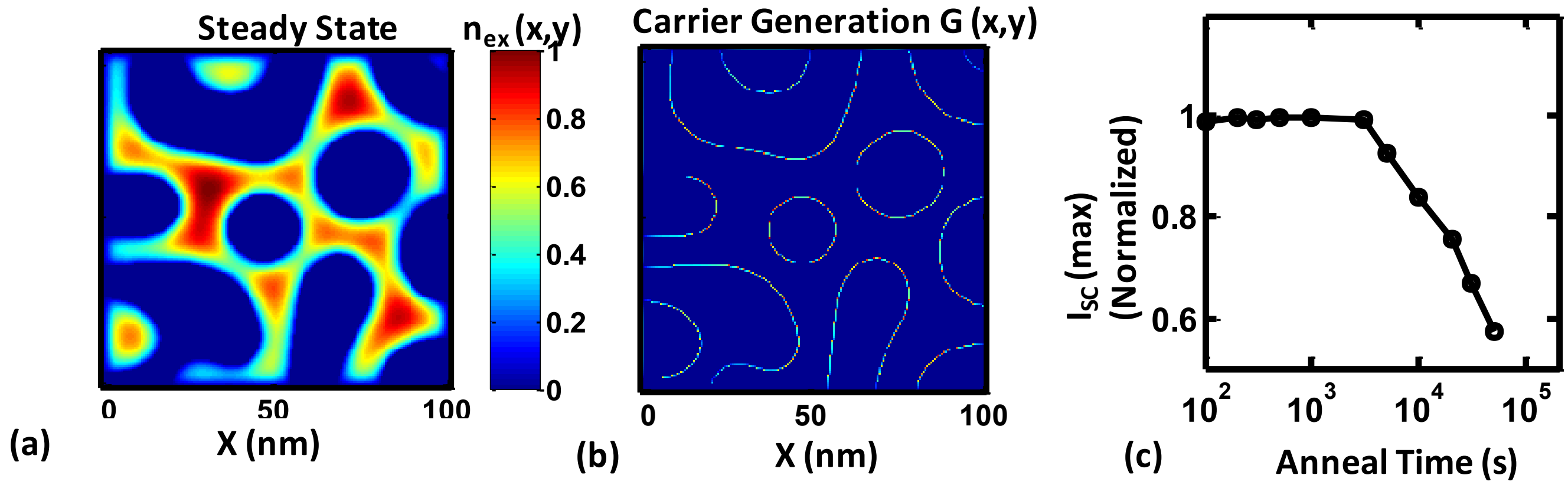

Figure 3: The gradual build up of exciton concentration $(n_{ex})$ in the phase-segregated morphology is shown in (a) by color-coding. Dark blue regions the figure are the acceptor phase where we assume no exciton is generated. (b) Spatially distributed electron/hole pair generation, $G_{e-h}(s)$, is shown by the bright spots. Since exciton dissociates only at the interfacial regions, charges carriers are generated only along those interfacial boundary lines. (c) The steady state value of $I_{SC}(max)$ is plotted against anneal time after normalizing the current by total exciton generation rate.

Usually, the field distribution inside the device is calculated by the transfer matrix method described by Pettersson[36]. In this approach, the materials inside the cell are characterized by the complex refractive indices and the interface between the two materials is modeled by the interface transfer matrix, which consists of complex reflection and transmission coefficients. The central quantity calculated in this approach is the point wise optical power dissipation of the electromagnetic field inside the cell as given by

$$Q(r) = \frac{1}{2} c\epsilon_0 \alpha\eta \left|E_{opt}(r)\right|^2 \quad (4)$$

Here, $c$ is the speed of light in vacuum, $\epsilon_0$ is the permittivity of vacuum, $\eta$ is the real index of refraction, $\alpha$ is the absorption coefficient, and $E_{opt}(r)$ is the total optical field at the point $r$. The number of photo-generated excitons is directly dependent on the optical power absorbed by the material, $Q(r)$, and hence exciton generation is proportional to $\left|E_{opt}(r)\right|^2$.

In the following discussion, we will proceed with the assumption of uniform exciton generation for simplicity, so that we can focus on the connection between morphology and transport. The model developed however is quite general and position-dependent exciton generation can easily be incorporated to define the overall performance of the solar cell.

### 3.2. Exciton Transport

When photons are absorbed in the active layer as described in section 3.1, they generate exciton in the absorbing polymer (P3HT). While the physics of exciton in polymer system remains an intriguingly complex theoretical issue[37], for the purpose of the present analysis we consider exciton as a charge neutral particle containing electron and hole under strong columbic attraction. Since excitons are empirically known to have a finite lifetime ($\tau_{ex} \approx 1$ ns), only a fraction of excitons so generated can reach the P3HT/fullerene distributed boundary (or charge separating zone) before being lost due to self-recombination. Unlike standard model in the literature where exciton diffusion is considered within a homogenized media, we explicitly consider the diffusion of the excitons towards the interfacial boundary by the continuity equation *within* the irreducibly complex 3D phase segregated geometry, defined by the interpenetrating network of donor/acceptor polymer layers, i.e.

$$\frac{\partial n_{ex}(r)}{\partial t} = D_{ex}\nabla^2 n_{ex}(r) - \frac{n_{ex}(r)}{\tau_{ex}} + G_{ex}(r) \ . \quad (5)$$

Here $n_{ex}(r)$ is the exciton density ($cm^{-3}$) at the point $r \equiv (x, y, z)$ of the active layer morphology, $D_{ex}$ is microscopic exciton diffusion constant independent of the morphology of the films, and $G_{ex}$ is the rate at which excitons are generated, which is assumed proportional to the photons absorbed (see section 3.1). We solve the exciton diffusion equation numerically by using the finite difference approach in 3D grid space of the phase-segregated geometry – thereby relating the exciton flux ($J_{ex}$) at the donor acceptor interface with the morphology of the cell.

The exciton concentration profile ($n_{ex}(r)$) on the phase-segregated geometry is obtained from the numerical solution of eq. (5). We use uniform exciton generation ($G_{ex}$) in the donor phase and no exciton generation in acceptor phase, as is typical for P3HT: PCBM system. The rate of charge carrier generation ($G_{e-h}(s)$), which takes place only on the interfacial nodes, is calculated from the exciton concentration profile ($n_{ex}(r)$). Fig. 3(b) shows a 2D plot of the spatially distributed generation rate of charge carriers. We numerically integrate this carrier generation rate at every grid point of the D/A interfacial surface throughout the volume of the cell to calculate the total current denoted as $I_{SC}(max)$. Note that this is the maximum limit of the short circuit current for a given morphology; the other loss mechanisms (e.g., geminate recombination, floating island, etc) are considered in Sec 3.3 and 3.4. The steady state values of $I_{SC}(max)$ are plotted in Fig. 3(c) as a function of anneal durations. We find that $I_{SC}(max)$ decreases rapidly for the morphologies corresponding to higher anneal times. This is mainly due to coarsening of the morphology beyond the exciton diffusion length.

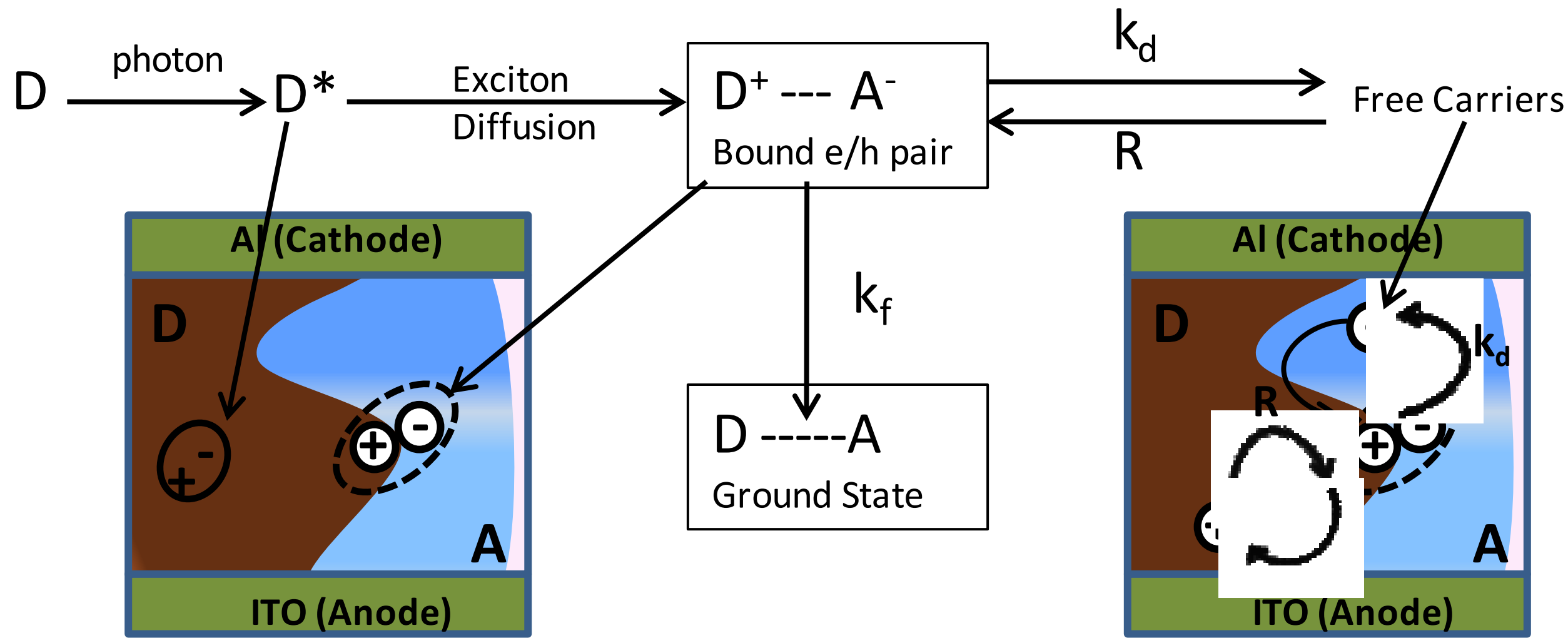

Figure 4: Schematic of the various events happening at the interface between donor (D) and acceptor (A). After the absorption of the photon in the donor polymer, exciton (D*) is generated. Exciton at the interface transfers its electron to the acceptor molecules, but the e/h pair remains bounded by the columbic attraction. The bound e/h pair dissociates into free carriers with a field dependent rate $k_d(E)$, or they decay to ground state with a rate $k_f$. Even the free carries can form again a bound e/h pair (with a rate R) if they come closer to the interface.

### 3.3. Charge Transfer/Separation

In the previous section, we calculated the fraction of the generated excitons that can reach the donor/acceptor interfacial region before being lost due to the self-recombination. Once the exciton reaches the D/A interface it transfers its electron from the donor side to the acceptor side and the rate of this charge transfer depends on the sharpness of the interface (details in supplementary material). After the charge transfer process, a geminate pair of a hole in the donor and an electron in the acceptor is formed. These electron hole pairs are strongly bound by Coulomb interaction (with binding energy in the range of 0.1-1 eV) because organic materials have low dielectric constants ($\epsilon_r \approx 3$).

These bound electron hole pairs must separate into free electrons and holes for the generation of photocurrent. The separation of the bound e/h pair depends on three distinct events, as shown in Fig. 4 (based on Ref[38] ). First, the geminate bound pairs of e/h either decay to the ground state with rate $k_f$, or (second) they are dissociated into free electrons and holes with the rate $k_d$. Interestingly, the free carriers close to the interfacial region can again form bound e/h pairs (third event) and the rate of this event is described by bimolecular recombination[38]. Combining the rates of all the three processes, the net carrier generation rate ($U$) is given by the following equation (detailed derivation given in Ref[38] ):

$$U(s) = P_d G_{e-h}(s) - (1 - P_d)\gamma\left(n_e(s)n_h(s) - n_{int}^2(s)\right). \quad (6)$$

Here, $s(x, y, z)$ stands for the points on the interfacial surface of the donor/acceptor region, $n_{int} = N_c \exp\left(-\frac{E_{gap}}{2kT}\right)$ is the intrinsic carrier concentration at the interface, $E_{gap} = LUMO_A - HOMO_D$ and the recombination strength $\gamma$ is given by Langevin[39], i.e $\gamma = q\langle\mu\rangle/\langle\epsilon\rangle$. $P_d$ is the field dependent exciton dissociation probability and $G_{e-h}(s)$ is the generation rate of bound e/h pair due to exciton dissociation as described in section 3.2. Note that unlike the classical implementation in homogenized media, the net carrier generation rate, $U(s)$, as well as the recombination rate, $R(s)$, depend on coordinates ($s \equiv (x, y, z)$) of interfacial surface within the active layer (and included as such in the numerical modeling) and therefore the carrier generation rate is morphology aware and resolved in position.

### 3.4. Electron and Hole Transport

Once bound e/h pairs are dissociated at the interface, the free electrons move in the acceptor phase and holes in the donor phase. To simulate the transport of electrons and holes in the disordered polymer network (without any effective media homogenization), we use drift diffusion model with self-consistent solution of the following set of equations:

Drift-Diffusion equation:

$$J_{e,h}(r) = q\mu_{e,h} n_{e,h}(r)(-\nabla V) \pm qD_{e,h}\nabla n_{e,h}(r) \quad (7)$$

Continuity equation:

$$\nabla J_{e,h} = \pm qU(s) \quad (8)$$

Poisson's equation:

$$\nabla(\epsilon_r \epsilon_0 \nabla V(r)) = -q(n_h(r) - n_e(r)) \quad (9)$$

Here $J$ is the carrier current density, $n$ is carrier density, $\mu$ is carrier mobility, $D$ is diffusivity, $V$ is the potential inside the device and $U$ is the net (free) carrier generation rate. The subscript 'h' stands for hole while subscript 'e' denotes electron. The positive sign in eq. (7) and (8) is for electron current and negative sign is for hole current. The boundary conditions for the set of equations at the electrodes are obtained from the equilibrium carrier concentration. We assume cathode is grounded and a positive voltage ($V_a$) is applied at anode; i.e., $V(z=0) = V_a$ *and* $V(z = T_{film}) = 0$. For all other open boundaries we assume zero flux boundary conditions i.e., $\nabla V = \nabla n_e = \nabla n_h = 0$.

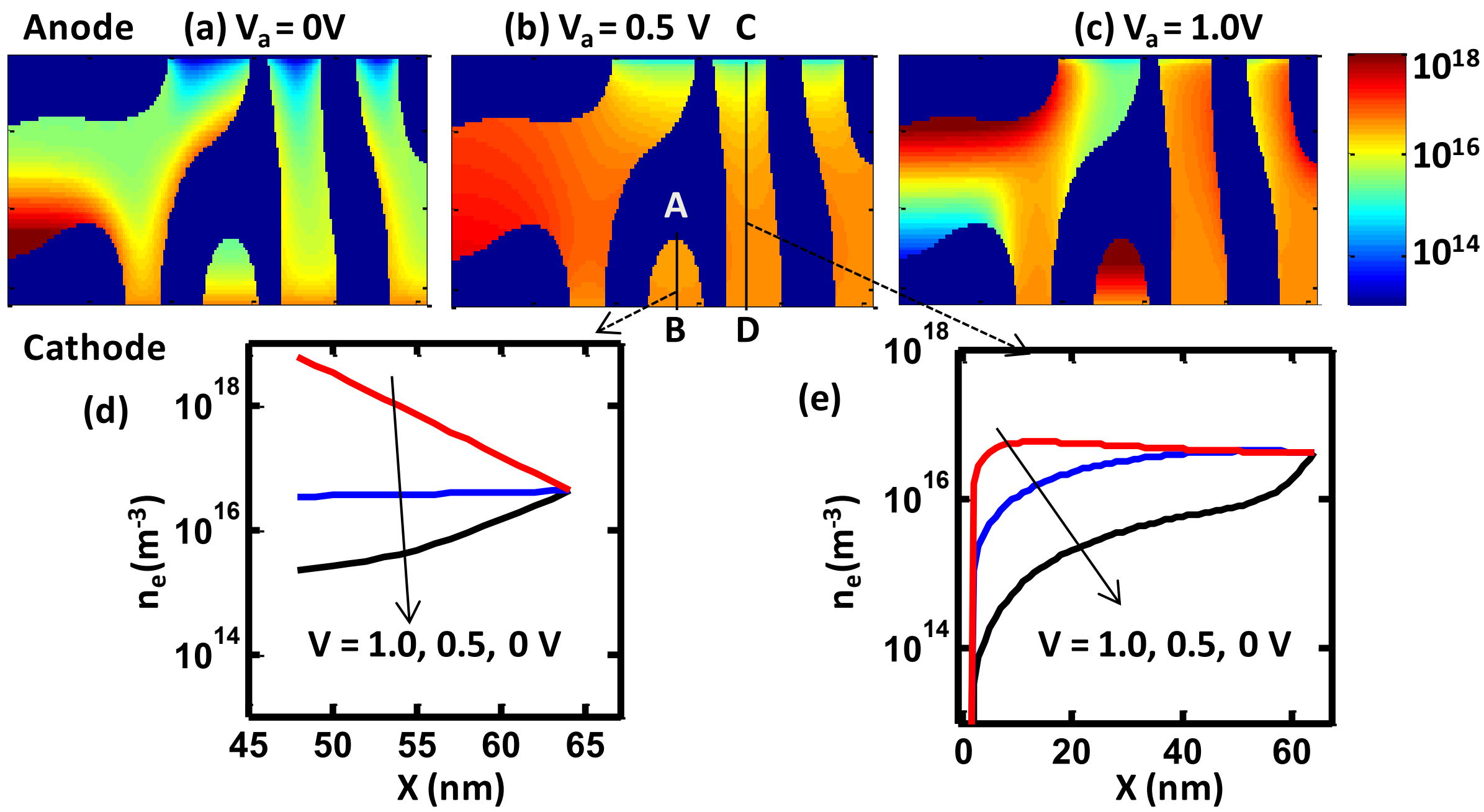

Figure 5: For a given morphology, carrier concentration profiles (a, b, c) are plotted for three different applied biases. (d) Electron concentration profile, along the vertical cut AB as shown in (b), is plotted for three applied biases. Note that the region where the AB cut is taken is connected to only one electrode. (e) Electron concentration profile along CD as shown in (b). CD represents the regions that are connected between both the electrodes.

To calculate the net current from the cell, the transport equations (eq. (7)-(9)) need to be solved in the interpenetrating complex polymer network corresponding to a particular set of processing conditions. Given the strong non-linearity, we use Scharfetter-Gummel's discretization scheme[40] for the drift diffusion equation in the 3D grid space. We find that the electric field in the device remains essentially constant between the electrodes (given by $E_z = (V_{bi} - V)/T_{film}$) because the polymer material is intrinsic and hence the solution of the Poisson's equation gives linear variation in potential. Note that in general the active layer may contain a large number of floating islands (also seen in Fig. 2(a-c)) which does not contribute to the steady state output current. Hence, once we simulate the active layer morphology, we identify the electrically connected percolating network and solve the carrier transport equations numerically only on those regions. Note that such morphology specific details cannot be accounted for in standard homogenized, one-dimensional models for BH solar cells.

## *4.* Results and Discussion

Given the simulation infrastructure, let us illustrate the capability of the simulation methodology by exploring the performance of the cell as a function of the process conditions. The uniqueness of the simulation framework is that it provides a mathematical description of the active layer morphology as a function of various process variables. Subsequently, the solution of the carrier transport equation on the corresponding morphology explicitly relates the performance of the solar cell to the underlying process conditions.

### 4.1. Current-Voltage characteristics of BH solar cell:

The current-voltage characteristics of the cell is obtained by the solving the transport equations (7-9) in the simulated active layer morphology. The numerical solution of these coupled equations (7-9) is shown in Fig. 5(a-c), where we plot the electron concentration profile for three different applied bias voltages. For better illustration, we also plot the electron concentration along the 1D cut (AB and CD) as shown in Fig. 5(d, e). Electron and hole concentration remain fixed to its equilibrium values (determined by the electrode work functions) at the electrode/semiconductor boundary. When the cell is forward biased, electron and hole concentrations increase gradually at the anode (ITO) and cathode, respectively, and current begins to flow through the percolating network. For an interesting validation of the model, let us focus not on the percolating network, but in regions *connected to a single electrode* (e.g., acceptor region connected to cathode but disconnected from anode as indicated by the AB cut in Fig. 5b). Since carriers in these regions cannot escape to the other contact, during dark conditions, the build-up of carriers is such that the diffusion flux exactly balances the drift-flux so that there is no current flow though such islands. With photo-generation however, the carrier density increases in the junction (i.e. the point A in Fig. 5b) and a net current flows out of the contact.

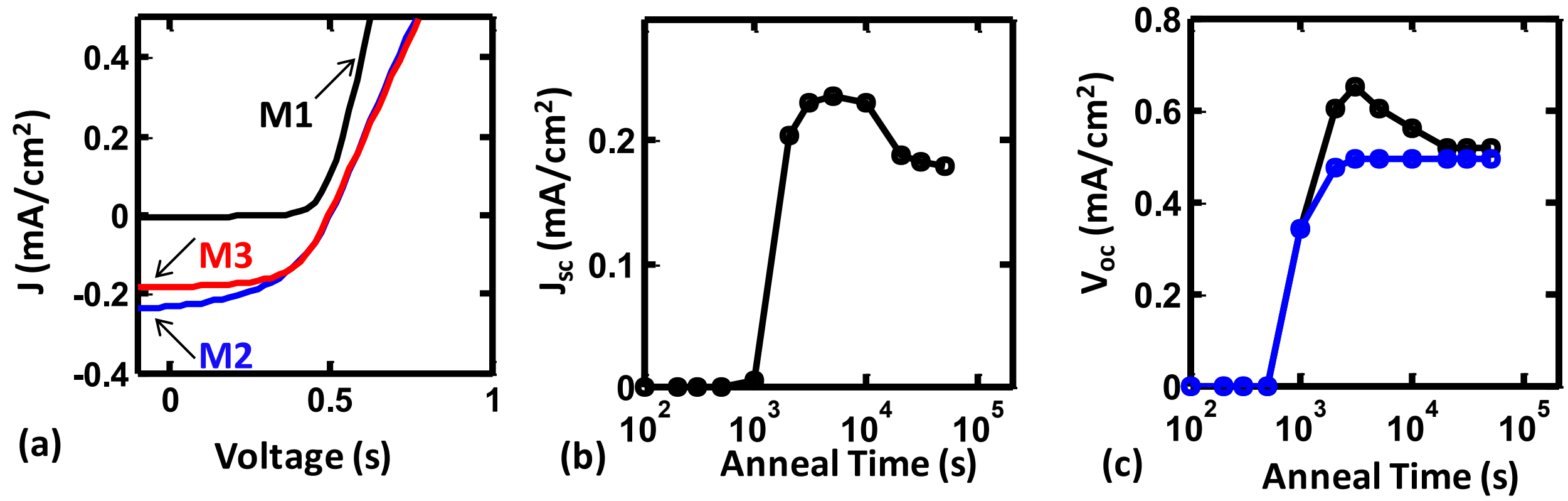

Figure 6**:** (a) The I-V characteristics obtained for the three morphologies (M1, M2, M3) shown in Fig. 2(a-c). (b) Anneal time dependent short circuit current is plotted for a number of different morphologies corresponding to different anneal duration. (c) Open circuit voltage variation with anneal time, with (blue) and without (black) interface recombination.

With such non-trivial distribution of charge carriers and flux pathways, it is not surprising that the I-V characteristics appear to behave rather anomalously as a function of process conditions. For example, let us explore the I-V characteristics of the cell for three different morphologies (M1, M2, M3 in Fig. 2(a-c)) corresponding to the three different anneal times ($t_a = 10^2, 10^3, 10^4\ s$) in Fig. 6(a). We note that both short circuit current and open circuit voltage vary significantly with the nature of phase-segregated morphology. To illustrate the variation of short circuit current as a function of anneal time, we plot the value of $J_{SC}$ obtained by numerical simulation on a number of morphologies corresponding to different anneal time in Fig. 6(b) and this plot clearly shows that there is a optimum anneal time which gives maximum short circuit current. Current is low at the initial phase of annealing because of the presence of floating islands in the active layer which cannot contribute to $J_{SC}$. $J_{SC}$ is also reduced at longer anneal times due to coarsening of the morphology which reduces exciton collection. Existence of such optimum anneal time has been observed empirically[7], but in this work we relate the optimum anneal time to the blend ratio, polymer chain-length, anneal temperature, etc.

Open circuit voltage ($V_{OC}$), on the other side, strongly depends on the interface recombination as shown in Fig 6(c). We find that if the interface recombination is weak, then $V_{OC}$ of the cell shows similar trends as shown by $J_{SC}$. This is mainly because dark current remains unchanged with annealing when interface recombination is weak and hence $V_{OC}$ follows the trends of $J_{SC}$ according to the equation: $V_{OC} = \frac{kT}{q} \ln\left(1 + \frac{J_{SC}}{J_0}\right)$. However, with strong interface recombination, dark current decreases with reduced interfacial area along with $J_{SC}$. Hence, $V_{OC}$ remains unchanged, even though net interfacial area reduces with higher annealing. While many of these features have been reported in literature[3,7], none of these morphology-sensitive features would be accessible to traditional homogenized, effective media theory of solar cells and therefore have never been explained satisfactorily.

### 4.2. Optimum annealing condition:

In the previous section we showed that there exists an optimum anneal duration for which $J_{SC}$ and $V_{OC}$ are both maximized. Although many empirical observations are there in literature for the existence of such optimum anneal time[6,7], but the dependence of optimum time on process variables has never been explicitly established. With the proposed process/device modeling framework we can now show that optimum anneal duration depends on the following relation:

$$t_a(opt) \sim max\ (t_{hj}\ , t_{prc})\ . \tag{10}$$

Here, $t_{hj}$ and $t_{prc}$ are constants determined by the process conditions. $t_{hj}$ is the heterojunction formation time, which represents the minimum anneal duration required for the formation of heterojunction with sufficiently strong quasi-electric field for exciton dissociation and irreversible charge transfer. $t_{prc}$ is the percolation time, representing the anneal duration required for the formation of percolating pathways necessary for the charge carrier transport. Below we explain how these two anneal time ( $t_{hj}$ and $t_{prc}$) depend on the process conditions and dictate performance of solar cell.

When an exciton reaches the interface of donor/acceptor region, it transfers its electron to the acceptor molecule. This charge transfer at the interface of the BH solar cell is an ultrafast process ($1/\tau_{ct} \approx 10^{13} s^{-1}$) as observed experimentally[41]. Unlike MBE or MOCVD growth films, however, at the early stages of phase segregation, the interface between the donor and acceptor phases is too diffused for efficient charge transfer. A simple derivation of the charge transfer probability as a function of interface width ($\langle W_{int} \rangle$) is discussed in supplementary material, but the essence of the analysis is the following: For efficient charge transfer, the effective width ($\langle W_{int} \rangle$) of the interface (refer to Fig. 2(d)) should be close to its minimum/saturation value. Fig. 2(e) shows the evolution of the interface as a function of the anneal time, and we find that only after a certain duration of annealing ($t_{hj}$), the interface width saturates to its minimum value ($\langle W_{int}(min)\rangle \sim 1nm$). Thus, $t_{hj}$, sets up a constraint on the minimum anneal time which is essential for sharp interface formation and efficient charge transfer.

The second requirement for annealing is the formation of percolating pathways for efficient charge carrier transport. At the initial phase of anneal duration, the active layer contains many floating islands, which cannot contribute to the output current. However, as we anneal the sample for longer duration, the floating islands grow bigger, connect themselves, and ultimately form percolating pathways from the anode to cathode. The process simulation shows that connected volume in the morphology increases with anneal time (see Fig. S2), demonstrating how longer annealing improves carrier transport. This constraint for carrier transport due to the effect of 'connected volume' sets another lower bound for the optimum anneal time, defined as the percolation time ($t_{prc}$).

### 4.3. Morphology and Reliability of OPV

The degradation mechanisms of OPV are complicated and poorly understood in the literature. In a review article[42], Krebs had described several possible degradation issues of OPV, which includes chemical degradation of electrode metals and polymer molecules in the presence of oxygen and water, photo-oxidation of polymers, thermal degradation due to morphological change, etc. Let us now consider the last degradation mechanism, i.e. thermal degradation due to morphological change, from the perspective of process-device correlation effects discussed in this paper.

For annealing the sample beyond this optimum time, the feature size of the phase-segregated morphology grows bigger than the exciton diffusion length and as a result exciton harvesting becomes poor and short circuit current begins to fall. This phenomenon has inherent implications on the lifetime of the BH solar cell. Since phase separation is a continuous process, even during the normal operating conditions of the cell the evolution of the active layer morphology continues to take place and consequently the performance of the cell degrades with operational time. Many empirical observations are available in the literature[4,5,42-45], which confirm this time dependent degradation process of the BH solar cell. Even in some recent papers[46], it has been reported that the cell performance improves with operational time and later it starts to degrade. This initial improvement in performance is the consequence of the annealing the samples for less than the optimum anneal time. The rate of the degradation process depends on the operational temperature, as the kinetics of the phase separation is a temperature-activated process. The process device simulation framework developed in this work is well equipped to address this degradation issue for this kind of BH solar cell.

### Conclusion

In summary, we have developed a conceptual and computational framework, which is capable to connect the process conditions to the ultimate device performance. Starting from process simulation, we have modeled and simulated each step of the cell operation, like photon absorption, exciton diffusion, charge separation and the charge carrier transport. This model highlights the possibility of complementing the state of the art empirical approach by predictive theoretical models of spinodal phase separation and coupled exciton/electron/hole transport for improved performance of solar cells. The physical description of each of the sub-processes can be made more accurate and sophisticated, but even this first attempt to connect morphology to performance clarifies a set of features that are inaccessible to previous modeling approaches. For example, our simulations clearly demonstrate that there exist an optimum anneal time for the short circuit current as well as open circuit voltage achievable from the BH solar cell and that optimum time is uniquely defined by two constraints- one related to sharp hetero junction formation ($t_{hj}$) and the other due to the percolating pathway formation($t_{prc}$). We find that many of the puzzling features related the shape of I-V characteristics (e.g insensitivity of $V_{OC}$ even with high interface recombination, sharp rise and fall of $J_{SC}$ with annealing) are simple consequences of the morphology of the active layer. Finally, based on the developed simulation framework, we discuss the possibility of predicting the intrinsic reliability of OPV.

### Acknowledgement.

The work was supported by the Center for Re-Defining Photovoltaic Efficiency through Molecule Scale Control, an Energy Frontier Research Center funded by the U.S. Department of Energy, Office of Science, and Office of Basic Energy Sciences under Award Number DE-SC0001085. The computational resources for this work were provided by the Network of Computational Nanotechnology under NSF Award EEC-0228390.